\documentclass[11pt]{article}
\usepackage{amsmath}
\usepackage{amsthm}
\usepackage{amssymb}
\usepackage{fullpage}
\usepackage{graphicx}

\begin{document}
\title{\textbf{Proton Transport Entropy Increase \\In Amorphous SiO\textsubscript{2}}}
\author{Randall T. Swimm\\ {\small NASA JPL Contractor (retired), Pasadena CA USA E-mail: rtswimm@well.com}}
\maketitle

\begin{abstract}This paper presents a classical thermodynamic calculation of a Greens function that describes the declining rate of entropy growth as protons move under an applied electric field, through an amorphous SiO\textsubscript{2} layer in a MOS field-effect device gate oxide. The analysis builds on work by McLean and Ausman (1977) and Brown and Saks (1991). Polynomial models of fitting parameters dB/d$\alpha$, y\textsubscript{0}, and A/y\textsubscript{0} based on interpolation TABLE I of McLean and Ausman are presented. Infinite boundary conditions are introduced for the parameter dB/d$\alpha$. Polynomial representations are shown of dB/d$\alpha$, y\textsubscript{0}, A/y\textsubscript{0} and the Greens function as a function of the dispersion parameter $\alpha$. The paper shows that parameters y\textsubscript{0} and A/y\textsubscript{0} are nearly conic sections with small residuals of a few percent. This work is intended as a first step toward a near-equilibrium thermodynamic continuous-time random walk (CTRW) model (anomalous diffusion) of damage introduced into thick-oxide silicon-based power\-MOS parts by space radiation effects such as those found in the Jovian radiation belts. Charge transport in amorphous silica electrical insulators is by thermally activated tunneling, not Brownian motion.
\end{abstract}

\section{Introduction}

\paragraph{The goal of this paper is to show how thermodynamics might be applied to the study of the dispersive transport of charges released by radiation damage of thick amorphous silica such as those found in power\-MOS parts in a space radiation environment. There are four aspects to this goal. The first, and possibly most important aspect is that dispersive transport of charges released by radiation damage is a thermally-activated tunneling phenomenon (Boesch, McLean, McGarrity,  and Ausman(1975)[1], Saks, Klein,  and Griscom(1988)[2], and Grasser(2012)[3]), and emphatically not a Brownian motion phenomenon, in that not only is the random timing of hops determined by the random nature of tunneling, not by random collisions of particles in a fluid, but also to tunnel is to arrive at the destination site immediately, rather than via a classical hydrodynamic transport process. The second is that radiation damage in thick oxides of high voltage MOS parts releases charge carriers in the oxide that when subject to dispersive transport under electric fields results in irreversible lattice relaxation (Prigogine, Mayn\'e and De Haan(1977)[4], and Friedrichs(1948)[5]), which would imply a corresponding increase in entropy. The third is that high-voltage thick-oxide silicon parts are known to be among the microelectronic parts most readily damaged at high total ionizing doses. The fourth is that thermodynamics has not been widely applied to radiation damage of power\-MOS parts, if at all. Finally, although this paper applies classical thermodynamics to the chosen example, a proper treatment would be to apply near-equilibrium thermodynamics (Metzler(2019)[6]) rather than classical thermodynamics.}

\section{Polynomials}

G(Z) = (A$\theta$/y\textsubscript{0})(1 + BZ)\textsuperscript{$\alpha$-1/2} e\textsuperscript{-Z} for Z \textgreater  0 and \\ G(Z) = 0 for \textless 0.

\paragraph{This work mainly builds on two prior works, that of F. Barry McLean and George A. Ausman, Jr.: Simple approximate solutions to continuous-time random-walk transport (1977)[7] and that of D. B. Brown and N. S. Saks: Time dependence of radiation-induced interface trap formation as a function of oxide thickness and applied field (1991)[8]. In equation (42) of McLean and Ausman, the authors present what they describe as the simplest trial function which satisfies the asymptotic requirements, a Greens function G(Z), as presented above, where $\alpha$ is the dispersion parameter, and B, y\textsubscript{0}, and A are adjustable parameters and where $\theta$ = (1-$\alpha$)\textsuperscript{-1} and Z = (y/y\textsubscript{0})\textsuperscript{$\theta$}. They then solve the Greens function by the method of moments to obtain three simultaneous coupled equations, which they state must be solved numerically. They then go on to propose interpolation  to provide approximate solutions to the moment equations. Trial function parameters for eleven values of $\alpha$ for the parameters A, B, and y\textsubscript{0} are tabulated in TABLE I.}

\paragraph{This author recommends that a preferred set of parameter values be calculated for B, y\textsubscript{0}, and A/y\textsubscript{0}. This is both because A appears in the Greens function only in the quotient A/y\textsubscript{0}, and also because polynomial solutions for B, y\textsubscript{0}, and A/y\textsubscript{0} can be derived across a broad range of $\alpha$ values, whereas the parameter A can be fit only with polynomial segments over short domains of $\alpha$. Moreover, y\textsubscript{0} and A/y\textsubscript{0} are both found to be quite close to conic sections with small residual deviations that can themselves be fit with polynomial approximations for improved precision. This paper will use the approach of fitting polynomials for the $\alpha$ values presented in TABLE I of McLean and Ausman, extended to include A/y\textsubscript{0} (Table I).}

\section{Parameter B}

(dB/d$\alpha$)*($\alpha$)*(1-$\alpha$) = -0.0125347 - 0.2345602$\alpha$\textsuperscript{1} + 4.3930191$\alpha$\textsuperscript{2} - 8.0764062$\alpha$\textsuperscript{3} + 7.0601918$\alpha$\textsuperscript{4} - 3.0268655$\alpha$\textsuperscript{5}

\paragraph{Brown and Saks present a solution for the parameter B in their Appendix A by defining B=10\textsuperscript{dB/d$\alpha$} developing a polynomial solution for the derivative dB/d$\alpha$. However, the parameter B in Table 1 of McLean and Ausman already contains a boundary value of infinity at $\alpha$ = 1, while dB/d$\alpha$ has an additional singular boundary value of minus infinity at $\alpha$ = 0. But Brown and Saks chose to ignore those two singularities in dB/d$\alpha$ and present a third degree polynomial approximation for dB/d$\alpha$.} 

\paragraph{In this paper, the author chose instead to remove both singularities by multiplying the dB/d$\alpha$ tabulation derived from Table 1 by the product $\alpha$($\alpha$-1), and then by fitting that well-behaved result by a polynomial, while imposing boundary values of 0 at $\alpha$=0 and $\alpha$=1. There are some subtleties since we are dealing with infinite boundary values. The full solution will involve multiplying and later dividing by the above product, which for all but the end points involves multiplying and dividing by non-zero numbers with a quotient of one. But for the end points we would be multiplying and dividing by zero. From among all the possible quotient values at the end points, we will choose the quotient value of 1, so as to avoid a discontinuity at the end points. We also note that we could use $\alpha$ raised to any power, and still obtain a quotient of 1 after multiplying and dividing. We will choose to raise $\alpha$ by the power 1, and so also for the factor ($\alpha$-1). Boundary limits cause max oscillation differences around \mbox{$\alpha$ =} 0.8, where dB/d$\alpha$ changes by 2\% from 2.4010 for Brown and Saks to 2.3515 for this model.}

\begin{figure}[h]
     \centering
     \includegraphics{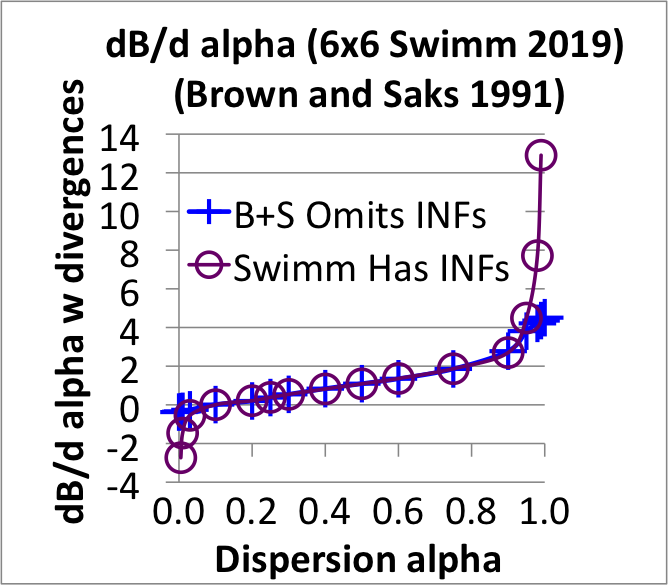}
     \caption{Slope of parameter B versus alpha in which Brown and Saks omit divergences}
     \label{6x6_B-slope_image}
 \end{figure}

\paragraph{The polynomial solution was obtained by seeking a solution on a 6x6 matrix using Excel, which corresponds to a degree 5 polynomial. The sampled values from McLean and Ausman Table I were chosen at $\alpha$ values of 0.1, 0.25, 0.4, 0.6, 0.75, 0.9, omitting 0.2, 0.3, and 0.5. The numerical analysis goal was an exact fit at the sampled Table I values, since the analysis was carried out by hand in Excel. This near-exact requirement on a subset of Table I values allowed the author to test for any blunders introduced. Nearly exactly fitting a subset of carefully chosen Table I values allowed the author to identify the lowest power polynomial that faithfully fit as many of the interpolation points as possible, while minimizing the intrinsic oscillations of each polynomial, known as the Gibbs phenomenon.} 

\section{The Conic-section parameters}

\paragraph{The plotted values of y\textsubscript{0} from TABLE I of McLean and Ausman are quite close to forming an open-downward parabola, to within 6\%. Also, the plotted values of A/y\textsubscript{0} calculated from our extended Table I of McLean and Ausman are quite close to the left half of an open-upward hyperbola, to within 2.5\%. This close agreement of y\textsubscript{0} and A/y\textsubscript{0} with conic sections offers the option of treating departures of the Table I values as perturbations from the conic section formulae. The recognition that two of the parameters of the extended Table I are quite close to being conic sections is a major simplification in the analysis of dispersive transport of charge carriers released by ionizing radiation in the amorphous silicon gates of power\-MOS devices. For the case of the hyperbola, the fit to the Table I entries is possible only for $\alpha$ values greater than or equal to $\alpha$=0.25.}

\subsection{Parameter y\textsubscript{0} Parabola}

y\textsubscript{0} = -(3.09016)($\alpha$-0.5)$\alpha$ + 1.77245

\paragraph{The parabola equation above roughly fits the Table I values for y\textsubscript{0} versus $\alpha$.}

\subsection{Parameter A/y\textsubscript{0} Hyperbola}

A/y\textsubscript{0} = {(1 + ((x+x\_offset)\textsuperscript{2} /b\textsuperscript{2}) * a\textsuperscript{2})\textsuperscript{1/2}} - A/y\_offset)

\paragraph{The hyperbola equation above roughly fits the derived Table I values for A/y\textsubscript{0}, where (y\_offset)\textsuperscript{2} = 1/3, x\_offset = -1, a = 1, b\textsuperscript{2} = 1/3, and x spans the $\alpha$ values 0 to 1, which upper limit is at the minimum of the upper hyperbola branch. x\_offset and y\_offset shift the hyperbola down and too the right from where it would be symmetric relative to the origin.}

\paragraph{Only B had been treated with polynomials until now. The power series terms for B and for the following conic-section residuals alternate in sign. This implies that the results involve differences that cancel many significant digits in the sum. Thus, many more significant digits must be retained in the polynomial sums in order that a sufficient number of significant digits survive after the differences are taken.}

\section{Conic-section residuals}

\paragraph{The parameters in the extended McLean and Ausman Table I are not exact conic sections. This section discusses polynomial representations of departures from the conic sections, to which this paper refers as residuals $\Delta$.}

\subsection{Parameter y\textsubscript{0} Parabola residual $\Delta$\textsubscript{p}}

$\Delta$\textsubscript{p} = 1.71647354$\alpha$ - 18.22930264$\alpha$\textsuperscript{2} + 88.97609928$\alpha$\textsuperscript{3} - 251.75918$\alpha$\textsuperscript{4}  + 433.42061495$\alpha$\textsuperscript{5} - 445.1258$\alpha$\textsuperscript{6} + 250.0156671$\alpha$\textsuperscript{7} - 59.01457068$\alpha$\textsuperscript{8}

\begin{figure}[h]
     \centering
     \includegraphics{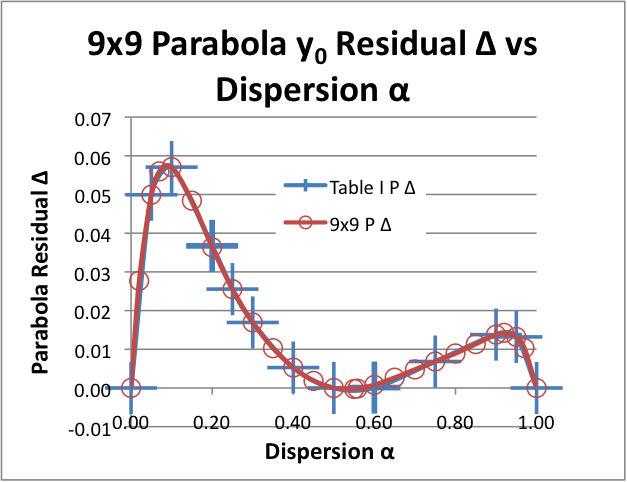}
     \caption{9x9 Matrix fit to residual of near-parabolic concave-down parameter y\textsubscript{0}. The two thick crosses are not included as fitting points.}
     \label{9x9_parabolic_residual_image}
 \end{figure}

\paragraph{The constant term in the $\Delta$\textsubscript{p} equation is zero. The polynomial for $\Delta$\textsubscript{p} initially did not satisfy the boundary conditions. In order to force the boundary values at $\alpha$=0 and $\alpha$=1, the author used a hack, introducing fictitious data at $\alpha$=0.05 and 0.95. He then adjusted the coupled fictitious entries at those $\alpha$ values by alternately adjusting the fictitious table entries until the boundary values were satisfied. Consequently, the results at the extreme low and high alpha values should not be used. However, it is unlikely that dispersion values at those extremes would be found for actual samples. The initial pioneering work by McLean and Ausman did indicate that $\alpha$ values less than 0.2 were measured. It appears that the early work of McLean and Ausman did not treat the possibility of more than one positive charge carrier, allowing only for holes in the radiation-damaged gate oxides. But Brown and Saks had realized that both holes and protons existed in radiation-damaged gate oxides. Protons move much more slowly than holes in amorphous silica. Therefore what had originally been interpreted as highly dispersive transport by a single charge carrier type consisting only of holes, was later recognized as being due to (at least) two charge carriers. The less dispersive $\alpha$ values would thus be closer to dispersion parameter values approaching $\alpha$ = 0.4. The sampled values from McLean and Ausman Table I were chosen at $\alpha$ values of 0.05 (adjusted), 0.10, 0.25, 0.30, 0.40. 0.50. 0.75, 0.90. and 0.95 (adjusted), omitting values 0.2 and 0.6.}

\subsection{Parameter A/y\textsubscript{0} Hyperbola residual $\Delta$\textsubscript{h}}

$\Delta$\textsubscript{h} = 0.0851632 -0.9667658$\alpha$ + 5.42831276$\alpha$\textsuperscript{2} - 16.85726381$\alpha$\textsuperscript{3} + 27.85164751$\alpha$\textsuperscript{4} - 22.67885949$\alpha$\textsuperscript{5} + 7.15009312$\alpha$\textsuperscript{6} 

\begin{figure}[h]
     \centering
     \includegraphics{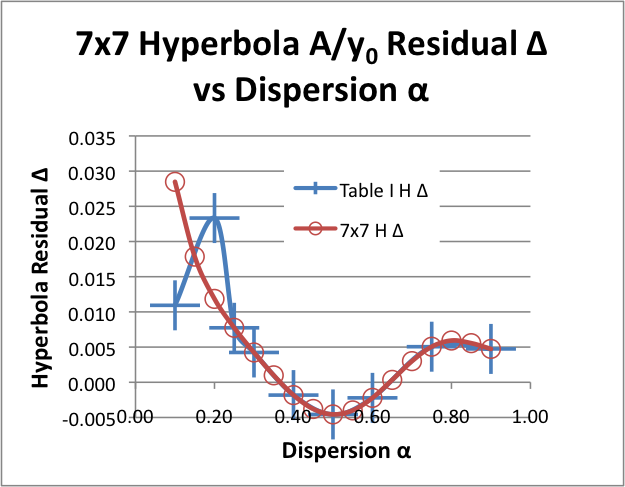}
     \caption{7x7 Matrix fit to residual of near-hyperbolic concave-up parameter A/y\textsubscript{0}. The residual model fails to fit the McLean and Ausman interpolation Table I for alpha values less than 0.25.}
     \label{7x7_hyperbolic_residual_image}
 \end{figure}

\paragraph{The hyperbola residual $\Delta$\textsubscript{h} was more difficult to treat as a polynomial. It was not possible to extend the polynomial fit below the $\alpha$ value 0.25. In a hyperbola residual plot, the fit to the higher values of $\alpha$, can be extrapolated to lower $\alpha$ values, to place the deviation of the lower $\alpha$ values in context. However, as discussed previously, values for dispersive transport $\alpha$ would be expected to fall well within the fitted portion of the interpolation points. The fitted sampled values from McLean and Ausman Table I were chosen at $\alpha$ values of 0.25, 0.30, 0.40, 0.50, 0.60, 0.75, and 0.90, omitting values 0.1, and 0.2.}

\section{Comments}

dS\textsubscript{p}/dt = V\textsubscript{g} * I\textsubscript{ox}/T

\paragraph{This paper is intended to point out that nonequilibrium thermodynamics might be brought to bear in the study of ionizing radiation damage to the gate oxides of silicon power\-MOS devices. In addition, this paper also attempts to provide polynomial approximations to model parameters that can be used to approximate the moment equations introduced by McLean and Ausman for similar purposes. The new polynomial models for parameters B, y\textsubscript{0} and A/y\textsubscript{0} constitute most of the discussion in this paper. This paper therefore augments the masterful work of Brown and Saks through the new polynomial treatments of y\textsubscript{0} and A/y\textsubscript{0} presented here. The author applies the word masterful to that work because the experiment was successfully designed to monitor only proton transport.}

\paragraph{An important part of the paper by Brown and Saks was their presentation of measurements of the accumulation of interface damage at a declining rate, following a brief ionizing irradiation of the part. As the interface damage increased at a declining rate, so it would also follow that the rate of entropy increase would be at a declining rate, as fewer and fewer of the released protons remained in transit as more of the protons were captured at interface traps. This relation is represented in the equation at the head of the Discussion section, where dS\textsubscript{p}/dt is the classical declining rate of proton transport entropy increase, excluding proton release by holes or proton chemical reactions, V\textsubscript{g} is the gate voltage, I\textsubscript{g} is the proton current in the gate oxide, and T is the absolute temperature, and no useful work is done by proton transport. The protons were released by holes that encountered passivating hydrogen atoms on non-bridging oxygen atoms, following the irradiation pulse. The declining proton current might be inferred from a measurement of the time evolution of the interface trap charge as measured by Brown and Saks.}

\paragraph{There are a few additional references that closely bear on the work of Brown and Saks. The work of Godet(2006)[9] showed that proton transport involves cross-ring tunneling between second-nearest neighbors. Proton tunneling distances were found by Godet to be 0.23 nm plus or minus 0.01 nm, which could be relevant to proton hopping distances, as the tunneling distance reported by Godet is much smaller than characteristic hopping distances reported by Brown and Saks. Johnston, Swimm, Thorbourn, Adell, and Rax(2014)[10]  addressed 3-D boundary conditions for micro-electronic parts that may also be relevant to any attempts to conduct electronic or calorimetric measurements on power\-MOS parts.}

\paragraph{The author thanks the JPL Library staff for their diligent efforts for over a year, to obtain journal articles during the background research phase, as the author learned about the field of nonequilibrium thermodynamics. The author is grateful for the interest of program managers and line managers during the research. The author thanks Laura for reviewing the paper with a type setter's eye. The author thanks Jean-Pierre Fleurial for agreeing to be the principal investigator, as contractors may not be a PI at JPL. Jean-Pierre and the author collaborated to obtain funding, but conducted their own research separately.}

\paragraph{The author considers this research to possibly constitute a new subfield (although other similar applications of thermodynamics exist). The author chose an open access journal in the hope that open access might allow physicists, chemists, engineers, mathematicians, and materials scientists an opportunity to collaborate in advancing the understanding of dispersive tunneling transport of charge carriers due to radiation damage in power-control-device amorphous SiO2 in the gate and other high-field insulating regions, in support of long-term presence and distant travel in space.}

\paragraph{What is more, in addition to this thermodynamic work possibly constituting a new subfield within radiation damage, it also represents a departure from the vast majority of works on dispersive charge transport, which are nearly all based on Brownian motion models. That is, dispersive transport by means of tunneling rather than by particle collisions itself constitutes a relatively unexplored subject outside the field of radiation damage of microelectronic part oxides. The author hopes that this, too, might attract researchers in a broad range of fields for this reason as well.}

\paragraph{This research was carried out at the Jet Propulsion Laboratory, California Institute of Technology, under a contract with the National Aeronautics and Space Administration and funded through the internal Research and Technology Development program during FY17-FY18. This research was also partially funded by JPL Section 5140 in FY19. \mbox{Randall} Swimm retired as a JPL contractor at the end of Fiscal Year 2019. Copyright California Institute of Technology 2019.}

\paragraph{ }
\bibliographystyle{plainnat}

\end{document}